\documentstyle[12pt,psfig]{article}
\def\beq{\begin{equation}}
\def\eeq{\end{equation}}
\def\bea{\begin{eqnarray}}
\def\eea{\end{eqnarray}}
\def\bq{\begin{quote}}
\def\eq{\end{quote}}

\def\gappeq{\mathrel{\rlap
{\raise.5ex\hbox{$>$}}
{\lower.5ex\hbox{$\sim$}}}}
\def\lappeq{\mathrel{\rlap{\raise.5ex\hbox{$<$}}
{\lower.5ex\hbox{$\sim$}}}}
\def\simlt{\stackrel{<}{{}_\sim}}
\def\simgt{\stackrel{>}{{}_\sim}}

\begin{document}
\pagestyle{empty}
\begin{flushright}
{CERN-TH/99-269\\
IFT-99/22\\
hep-ph/9910231}
\end{flushright}
\vspace*{5mm}
\begin{center}
{\bf FIXED POINTS IN THE EVOLUTION OF NEUTRINO MIXINGS}
\\
\vspace*{1cm} 
Piotr H. Chankowski$^{a,b)}$, Wojciech Kr\'olikowski$^{b)}$
and Stefan Pokorski$^{b)}$
\\
\vspace*{1.7cm}  
{\bf ABSTRACT} \\
\end{center}
\vspace*{5mm}
\noindent
We derive the renormalization group equations for the neutrino masses and 
mixing angles in explicit form and discuss the possible classes of their
solutions. We identify fixed points in the equations for mixing angles, 
which can be reached during the evolution for several mass patterns and
give $\sin^22\theta_{sol} =\sin^22\theta_{atm}\sin^2\theta_3/
(\sin^2\theta_{atm}\cos^2\theta_3 + \sin^2\theta_3)^2$,
consistently with the present experimental information. Further experimental 
test of this relation is of crucial interest. Moreover, we discuss 
the stability of quantum corrections to neutrino mass squared differences.
Several interesting mass patterns show stability in the presence of fixed point
solutions for the angles. 
\vspace*{1.0cm} 
\noindent

\rule[.1in]{16.5cm}{.002in}
\noindent
$^{a)}$ Theory Division, CERN, Geneva, Switzerland \\
$^{b)}$ Institute of Theoretical Physics, Warsaw University, Poland \\
\vspace*{0.2cm}

\begin{flushleft} 
CERN-TH/99-269\\
IFT-99/22\\
October 1999
\end{flushleft}
\vfill\eject
\newpage

\setcounter{page}{1}
\pagestyle{plain}

\section{Introduction}

Observation of   atmospheric \cite{ATMOSPH}
and  solar  \cite{SOLAR} neutrinos provide   important
indications    that neutrinos    oscillate   between  different   mass
eigenstates. Interpreting  each  observation in  terms of  two-flavour
mixing the oscillations  of   atmospheric  neutrinos  require  $\Delta
m^2_{atm}\sim2\times10^{-3}$ eV$^2$  and a nearly maximal mixing angle
$\sin^22\theta_{atm}\geq0.82$. For solar neutrinos, several  solutions
of   their deficit problem are possible.   The vacuum oscillation (VO)
solution     requires   $\Delta   m^2_{sol}\sim10^{-10}$  eV$^2$   and
$\sin^22\theta_{sol}\geq0.67$,  whereas the the so-called MSW solution
requires   $\Delta    m^2_{sol}\sim{\cal  O}(10^{-5})$   eV$^2$     and
$\sin^22\theta_{sol}\geq0.5$ (large  mixing angle solution - LAMSW) or
$4\times10^{-3}\simlt\sin^22\theta_{sol}\simlt1.3\times10^{-2}$ (small
mixing  angle   solution  -  SAMSW).    Combining these   results with
non-observation of  the  dissappearance of $\bar\nu_e$  in the reactor
experiments (in particular in CHOOZ \cite{CHOOZ})  
sensitive to $\Delta m^2\simgt10^{-3}$ eV$^2$
and $\sin^22\theta\simgt0.2$,   the   atmospheric and   solar neutrino
oscillations can be  explained  in terms of   three known flavours  of
neutrinos\footnote{We leave out the not yet confirmed LSND result
\cite{LSND}. All  three  measurements can   be
explained only if    there  exists  sterile neutrinos,     e.g.  light
right-handed ones  \cite{KR}. }, the  assumption which we adopt in this
paper,   as $\nu_\mu\rightarrow\nu_\tau$  and  $\nu_e\rightarrow\nu_x$
oscillations with $\nu_x=\nu_\mu$ if the mixing was two flavour one.

With accumulating data, considerable
attention  has  been  focused on   the    determination of  the   full
Maki-Nakagawa-Sakata    (MNS)    $3\times3$   unitary mixing    matrix
\cite{MANASA} which is conveniently  parametrized by 3 angles 
(we set the phase $\delta=0$)
\begin{eqnarray} 
U =\left(\matrix{ c_1c_3 & s_1c_3 & s_3 \cr
-s_1c_2-c_1s_2s_3 & c_1c_2-s_1s_2s_3 & s_2c_3 \cr
s_1s_2-c_1c_2s_3 & -c_1s_2-s_1c_2s_3 & c_2c_3}\right)
\label{eqn:mns}
\end{eqnarray}
where  $s_i$   and  $c_i$ denote   $\sin\theta_i$ and  $\cos\theta_i$,
respectively.

Defining $\Delta  m^2\equiv  m^2_2-m^2_1$ with $|\Delta m^2|=\Delta m^2_{sol}$ 
and $\Delta M^2\equiv m^2_3-m^2_2$
with $|\Delta M^2|=\Delta m^2_{atm}$, the CHOOZ experiment measures
(with $\Delta M^2\gg\Delta m^2$ and $\sin^2( \Delta m^2L/4E)\approx0$)
\begin{eqnarray} 
P(\bar\nu_e\rightarrow\bar\nu_e)=1 - 4 U_{13}^2(1-U_{13}^2)
\sin^2\left({\Delta M^2 L\over4E}\right)
\end{eqnarray}
and the negative   result   of the search    constraints $s_3=U_{13}$:
$s_3^2\simlt0.2$.

For the solar neutrino oscillations we have (averaging $\sin^2(\Delta
M^2 L/4E)$ to $1/2$):
\begin{eqnarray} 
P_{3\times3}(\nu_e\rightarrow\nu_e) = 
c_3^4P_{2\times2}(\nu_e\rightarrow\nu_e) + s^4_3
\end{eqnarray}
(for this formula to  account  also for the   MSW solution one  has to
mutiply the  electron density in the Sun  in the 2-flavour probability
$P_{2\times2}$ by   $c^2_3$ \cite{BIGIGR}).  In   the 
limit $s_3=0$, $\theta_{sol}=\theta_1$.   However, since in general    
$U_{31}\neq0$, $P(\nu_e\rightarrow\nu_\tau)\neq0$.

For the atmospheric neutrinos 
\begin{eqnarray} 
P(\nu_\mu\rightarrow\nu_\tau) = 4 (U_{23} U_{33})^2 
\sin^2\left({\Delta M^2 L\over4E}\right)
= 4 s^2_2c^2_2 c^2_3\sin^2\left({\Delta M^2 L\over4E}\right)
\end{eqnarray}
so, in  the limit $s_3=0$, $\theta_{atm}=\theta_2$.

Thus, up to some  uncertainties in $|U_{13}|=|s_3|$, for each mentioned
earlier solution to the atmospheric and solar neutrino problems we can
infer  the {\sl gross} pattern   of the $3\times3$ mixing matrix.  The
striking features  are large mixing angles  $\theta_2$ and, for VO and
LAMSW solutions, $\theta_1$. Measuring $s_3$ is very important, as the
dependence  of  the fitted  $s_1$ and $s_2$  on $s_3$   is not totally
negligible \cite{BAR,FOLIMASC,KIKI}. 

Large number  of papers    addressed the issue of
theoretical explanation    of  neutrino  masses  and  mixings   and of
incorporating it  into some global  solution  to the flavour  problem   
\cite{DIHARA,BLRATO,LELORO,LELOSCH,LOVE,BAHAST,ALFE,LORO,BAHAKARO,JESU,CAELLOWA,ST}. 
One attractive possibility for  giving  the  three flavours  of  neutrinos
small Majorana masses  is the seesaw mechanism  which generates in the
effective low energy theory (MSSM or the SM) a dimension five operator.
In the two-component notation it reads: 
\begin{eqnarray} 
{\cal L}_{dim~5} = - {1\over4M} {\cal C}^{ab} (Hl^a)(H l^b) + h.c.
\label{eqn:dim5op}
\end{eqnarray}
where  $a$ is   generation index, $l$  is the ordinary lepton $SU(2)$
doublet,  $H$ is the  hypercharge $+1/2$ Higgs doublet (i.e. $H^{(2)}$
in the  MSSM) and the matrix  ${\cal  C}$ is  dimensionless because we
have factorized out the  overall scale $M$  of the heavy, right-handed
neutrino mass matrix.  This  operator, after the  electroweak symmetry
breaking, is the origin of the left-handed neutrino Majorana mass term 
\begin{eqnarray} 
{\cal L}_{\nu ~mass} = - {1\over2} m_\nu^{ab} \nu^a\nu^b + h.c. 
\end{eqnarray}
with $m_\nu = (v^2/4M) {\cal C}$ where $v/\sqrt2=\langle H_2\rangle$. 
After  the     unitary  rotation  $\nu^a\rightarrow     V^{ab}_L\nu^b$
diagonalizing the neutrino\footnote{We will  always assume that $m_3$ 
is positive and  allow  for  negative signs  of  $m_1$  and/or $m_2$.} 
mass matrix ${\cal C}$
\begin{eqnarray} 
V_L^T {\cal C} V_L = C_D \equiv {\rm diag}(C_1,C_2,C_3)
\sim{\rm diag}(m_1,m_2,m_3)
\label{eqn:diagC}
\end{eqnarray}
the  charged current   weak neutrino interactions  depend   on the MNS
unitary matrix (\ref{eqn:mns}),  $U=E_L^\dagger V_L$ and the rotations
$E_L$ are defined by
\begin{eqnarray} 
E_L^\dagger{\cal M}^2_e E_L = H^2_e\equiv {\rm diag}(h^2_e,h^2_\mu,h^2_\tau)
\label{eqn:diagMe}
\end{eqnarray}
(${\cal M}^2_e\equiv Y^\dagger_eY_e$  is the square of the lepton
Yukawa coupling  matrix and $h^2_a$  are its eigenvalues).  
It is this  matrix
whose elements are probed in the neutrino oscilation experiments. 
In the effective theory below the
scale $M$, with the right-handed   neutrinos decoupled we can work  in
the     charged   lepton  mass eigenstate    basis,      fixed by the
diagonalization of the charged lepton Yukawa  matrix at the scale $M$
and then $U=V_L$.

The   measured neutrino    mass matrix  $m_\nu =
(v^2/M)UC_D   U^T$  is  linked to   the   fundamental  mass generation
mechanism by two steps.  The first one  (in the top-down  approach) is
the  renormalization group (RG) evolution  of the neutrino and charged
lepton Yukawa coupling matrices $Y_\nu$  and $Y_e$, respectively, from
the  scale   at which  they  are   fixed  by some   theory  of flavour
(presumably at the GUT  scale) down to  the right-handed Majorana mass
scale $M<M_{GUT}$.  At that scale  the right-handed neutrinos with the
Majorana  mass matrix $\hat M$ are  integrated out and one obtains the
Standard  Model (SM) or its  supersymmetric  extension (MSSM) with the
Majorana  mass matrix  of the  light neutrinos  given by  the operator
(\ref{eqn:dim5op}). 
The effective neutrino Majorana mass matrix ${\cal C}(M)$ at the scale $M$ 
is dependent on both, neutrino and charged lepton original Yukawa
textures at $M_{GUT}$ and on $\hat M$.  

The second step consists of the renormalization group evolution of this
operator   down  to  the     electroweak scale  $M_Z$,  which provides
unambiguous mapping of the pattern at  the scale $M$ into the measured
pattern at $M_Z$ scale. 
Once we choose to  work with the  SM  or the MSSM the  physics
below the scale $M$ is almost  unambiguous\footnote{In the of the
SM there is a  slight dependence of the  overall scale of the neutrino
masses on  the Higgs  boson quartic coupling   $\lambda$ (i.e. on  the
Higgs boson mass) and in the MSSM  on the assumed scale of 
supersymmetry breaking  which we will ignore  in the  discussion which
follows.}  and it is interesting to study the effects of quantum 
corrections to the neutrino mass matrix ${\cal C}(M)$ summarized in its 
renormalization group (RG) evolution down to $M_Z$. We
find it convenient to    work directly  with $m_a$'s and    $U_{ab}$'s
\footnote{For similar approach  in the quark sector see  \cite{OLPO}.}
rather  than with elements of the matrix $m_\nu$  since, as we shall see,
the evolution  of mass eigenstates and mixing  angles often allows for
easy qualitative discussion. 

In the present paper we extend earlier discussions  
\cite{BAROST,CAESIBNA_VO,HASU} by, first, writting
down   the  renormalization group   equations  for  the  neutrino mass
eigenvalues and mixing  angles between the three  flavours of neutrinos,  
in the MSSM (and the SM). We point out the main  differences between the 
evolution with $3\times3$  mixing  or $2\times2$  mixing. Next,
we discuss  in some detail  the  solutions to the  RGE's for mixing  
angles and their  dependence on the mass  eigenvalue patterns:
$m^2_3\approx\Delta M^2\gg  m_2^2, m_1^2$ (hierarchical), 
$m_1^2\approx m_2^2\approx\Delta
M^2\gg  m_3^2$ (inversely hierarchical) and $m^2_1\approx m^2_2\approx
m^2_3\sim{\cal O}(\Delta  M^2)$ or larger (degenerate),  which are   
consistent with the
measured mass squared differences. We identify several infrared (IR)
fixed points in the RG equations for mixing angles,
which can be reached for some range of neutrino masses and are compatible
with the present experimental information on the mixing angles.

Finally, we discuss the stability with  respect to quantum corrections of
the  three  neutrino   mass  eigenvalue patterns which turns out to be 
correlated with the character of the evolution of the angles. 
In our formalism, we can reproduce in a very simple way the earlier 
results \cite{BAROST,CAESIBNA_VO,HASU}, in particular,
those on the possibility of the VO  solution with the stability of the
inversely  hierarchical or degenerate  patterns. 

\section{RGE for neutrino masses and mixing angles}

The importance  of the scale dependence of  the coefficient ${\cal C}$
of the  operator  (\ref{eqn:dim5op})  have been  first   emphasized in
ref. \cite{WE,CHPL}.  The derived there renormalization  group  equation
in the  MSSM   reads\footnote{The MSSM RGEs
given  in ref.   \cite{CHPL} allow to  treat   also the case  in which
squarks and/or gluino  are  much heavier than sleptons,  charginos and
neutralinos  so that the    decoupling  procedure  \cite{CH}  can   be
employed; in this case there are four  different (in component fields)
operators which mix below the squark/gluino treshold. Above it one has
(in    the            notation    of        ref.          \cite{CHPL})
$c_1^{ab}=2c_{12}^{ab}=2c_{21}^{ab}=c_3^{ab}\equiv{\cal  C}^{ab}$  and
the four equations of ref. \cite{CHPL} merge into  the one quoted here
(derived independently also in ref. \cite{BALEPA}).}
\begin{eqnarray} 
{d\over dt} {\cal C} = -K {\cal C} -{\cal C} {\cal M}^2_e - ({\cal M}^2_e)^T
                     {\cal C} 
\label{eqn:rgeC}
\end{eqnarray}
with ${\cal M}^2_e=H^2_e$ (diagonal) in the basis we are working and where 
\begin{eqnarray} 
K\equiv\left[-{6\over5}g^2_1-6g^2_2+6{\rm Tr}{\cal M}^2_u\right], 
\end{eqnarray}
${\cal M}^2_u\equiv Y^\dagger_uY_u$ is  the square of the up-type quarks
Yukawa coupling  matrix,           $g_1^2\equiv(5/3)g^2_y$,
$t=(1/16\pi^2)\log(M/Q)$ and $M$ is the large Majorana scale.  
In the SM the correct form of
the equation  was given in ref.   \cite{BALEPA} and has  the same form
with   $-(6/5)g^2_1\rightarrow+2\lambda$ ($\lambda$ is the scalar quartic
coupling), $-6g^2_2\rightarrow-3g^2_2$, 6Tr${\cal M}^2_u\rightarrow 
{\rm Tr}(6{\cal M}^2_u+6{\cal M}^2_d+2{\cal  M}^2_e)$ and  the
coefficients of  the last  two  terms  are  $+1/2$  (instead of $-1$). 

Equation (\ref{eqn:rgeC}) can  be elegantly solved \cite{ELLO},  since
in  the absence  of right-handed   neutrinos below  the  scale $M$ the
matrix $E_L$ defined in (\ref{eqn:diagMe}) does not run. 
Thus, in the basis in which the leptonic Yukawa matrix is diagonal the
equation (\ref{eqn:rgeC}) has the obvious solution 
\begin{eqnarray}
{\cal C}(t) = I_K {\cal I} {\cal C}(0) {\cal I} 
\label{eqn:fullsol}
\end{eqnarray}
where ${\cal I}={\rm diag}(I_e,I_\mu,I_\tau)$, and 
\begin{eqnarray}
I_K\equiv\exp\left(-\int_0^t K(t^\prime) dt^\prime\right), ~~~~
I_a \equiv\exp\left(-\int_0^t h^2_a(t^\prime) dt^\prime\right).
\label{eqn:kfact}
\end{eqnarray}
The only  role of the factor $I_K$  is to change  the overall scale of
the   neutrino masses during the   evolution.  For  $M=10^{10-15}$ GeV, 
$I_K\approx0.9-0.6$ with smaller values   for lower  $\tan\beta$ 
due to the enhancement of the top quark Yukawa coupling in $K$.

Although  the solution to the RG equation for the  matrix 
${\cal  C}$  is  simple,
qualitative  features of the running of  the mass egenvalues $m_a$ and
the  MNS mixing  matrix $U$ are   often masked by the  diagonalization
procedure.  In this paper we derive the RGE directly for $m_a$ and $U$.

Following the  method  of ref.  \cite{BA},  the RG equation for the matrix
$V_L=U$ (in the charged lepton mass eigenstate basis) can be written as 
\begin{eqnarray}
{d\over    dt}U  =   -U    \varepsilon_\nu  ~~~~{\rm     with} ~~~
\varepsilon^\dagger_\nu =-\varepsilon_\nu 
\end{eqnarray}
where the  matrix  $\varepsilon_\nu$ is antihermitean, to preserve
unitarity of $U$. It is determined   by the  requirement   that 
 the matrix ${\cal C}(t)$ is diagonalized by $U(t)$
at any scale and the RHS of 
\begin{eqnarray}
{d\over dt}(U^T {\cal C} U)  = {d\over dt}C_D = -\varepsilon^T_\nu
C_D - C_D \varepsilon_\nu - K C_D - C_D  U^\dagger H^2_e U - U^T H^2_e
U^\star C_D 
\end{eqnarray}
is diagonal.  For real matrices ${\cal C}$ and ${\cal M}^2_e$
which we consider here, the MNS  matrix  $U$ is real   and
orthogonal and $\varepsilon_\nu$ is antisymmetric. We get therefore
\begin{eqnarray}
\varepsilon_\nu^{ab}&=& -{C_a+C_b\over C_a-C_b} (U^T H^2_e U)_{ab} ~~~~~
{\rm for} ~a\neq b,  ~~~~~\varepsilon_\nu^{aa}= 0.
\end{eqnarray}
The running of the eigenvalues is then given by 
\begin{eqnarray}
{d\over dt}C_a = -\sum_{b=e,\mu\tau}(K + 2 h^2_b U^2_{ba}) C_a, ~~~a=1,2,3
\label{eqn:runC}
\end{eqnarray}
and the running of elements of the MNS matrix is given by 
\begin{eqnarray}
{d\over   dt}U_{ab} = \sum_{d\neq   b}{C_d+C_b\over  C_d-C_b}  U_{ad}
(U^TH^2_eU)_{db}.
\label{eqn:runU}
\end{eqnarray}
If two of the three eigenvalues,
say $C_a$ and $C_b$, are  equal at some  scale $t$ there is the obvious
freedom  in   choosing   the  matrix $U(t)$,   corresponding   to the
redefinition   $U(t)\rightarrow\tilde U(t)=   U(t)R$  where  $R$ is  a
rotation in the $ab$ plane. This freedom is eliminated by
quantum corrections, since for the evolution,  $R$ has to be
fixed by the condition 
\begin{eqnarray}
(\tilde U^TH^2_e\tilde U)_{ab}(t) = (\tilde U^TH^2_e\tilde U)_{ba}(t)=0 
\label{eqn:crosscond}
\end{eqnarray}
so that  equation (\ref{eqn:runU}) in nonsingular. 

Eqs. (\ref{eqn:runC},\ref{eqn:runU})  give directly the running of the
the measurable parameters.      It is straighforward   to   derive the
equations  for  the three   independent  parameters  $s_1$, $s_2$  and
$s_3$. Neglecting $h_e$ and $h_\mu$ Yukawa couplings we get: 
\begin{eqnarray}
\dot s_1&=&
 - c_1(s_1s_2-c_1c_2s_3)(-c_1s_2-s_1c_2s_3)A_{21}h^2_\tau\label{eqn:runs1}\\
&-&s_1c_1c_2s_3(s_1s_2-c_1c_2s_3)A_{31}h^2_\tau
+c_1^2c_2s_3(-c_1s_2-s_1c_2s_3)A_{32}h^2_\tau\nonumber 
\end{eqnarray}
\begin{eqnarray}
\dot s_2&=& s_1c_2^2(s_1s_2-c_1c_2s_3)A_{31}h^2_\tau
-   c_1c_2^2(-c_1s_2-s_1c_2s_3)A_{32}h^2_\tau\phantom{aaa}\label{eqn:runs2}
\end{eqnarray}
\begin{eqnarray}
\dot s_3&=&-c_1c_2c_3^2(s_1s_2-c_1c_2s_3)A_{31}h^2_\tau
        -s_1c_2c_3^2(-c_1s_2-s_1c_2s_3)A_{32}h^2_\tau \label{eqn:runs3}
\end{eqnarray}
where $A_{ab}\equiv (C_a+C_b)/(C_a-C_b)=(m_a+m_b)/(m_a-m_b)$. 

Eqs.~(\ref{eqn:runC}-\ref{eqn:runs3})  give   us   several   immediate
results.  Neglecting  the  electron  and  muon  Yukawa  couplings, the 
solutions for squared mass eigenvalues read: 
\begin{eqnarray}
m^2_a(t)  =  m^2_a(0)I^2_K\exp\left(-\int_0^t4h^2_\tau(t^\prime)
U^2_{3a}(t^\prime)dt^\prime\right) 
\label{eqn:solmass}
\end{eqnarray}
Observe that, since $I^2_K<1$, the masses always decrease top-down. We also
see that the possibility of some change caused by the evolution in 
mass pattern resides solely in the differences in the
mixing matrix elements $U_{3a}$ and their RG running.  With
$h^2_\tau\approx(\tan\beta/100)^2$, $t\approx0.12$ for $M=10^{10}$ GeV
and $U^2_{3a}$ typically varying between 0 and 1/4 (except for
$U^2_{33}$) the exponent is at most of order of
$\epsilon\equiv h^2_\tau\log(M/M_Z)/16\pi^2\approx 
\tan^2\beta \times10^{-5}<2.5 \times
10^{-2}$ for $\tan\beta<50$.  We can then estimate the changes
in the mass squared differences: 
\begin{eqnarray}
\Delta m^2_{ab}(t)\equiv m^2_a(t)-m^2_b(t)=\Delta m^2_{ab}(0)    
- (\eta_a m^2_a(0)-\eta_b m^2_b(0))\epsilon 
\label{eqn:runmassdiff}
\end{eqnarray}
where we have neglected $I_K$ which is always close to 1 and the factors 
$\eta_{a(b)}>0$ are typically in the range $0-2$,
depending on the values of $U_{3a}$ factors and their evolution.
Taking (for definiteness) $\Delta m^2_{ab}(0)=0$, we see that the 
evolution of $\Delta m^2_{ab}(t)$ is limited by $m^2_a(0)\epsilon$
or $m^2_b(0)\epsilon$, i.e. by the value of the larger mass.

A spectacular feature of eq.~(\ref{eqn:runU}) or
eqs.~(\ref{eqn:runs1}-\ref{eqn:runs3}) is the existence of fixed points
at $U_{ab}=0$. The character of the fixed points (UV or IR) and the 
rate of approaching them depend on the mass factors $A_{ab}$. This point 
will be discussed in detail later on. We note
the possibility of a large
factor $A_{ab}$, for nearly degenerate and of the same sign
eigenvalues $m_a$ and $m_b$, which can permit a rapid approach to fixed 
points and drastically change the pattern
of mixing  at $M_Z$ during  the evolution from $M$.
\footnote{The possibility of strong changes in the mixing pattern due to
the evolution has been pointed out earlier for $2\times2$ mixing
\cite{BALEPA}.} In the solutions (\ref{eqn:fullsol}) for the matrix 
${\cal C}$ this point is encoded in the diagonalization procedure one has to
perform {\it after} the running. The equivalence of the two approaches
can be proven by diagonalizing the matrix ${\cal C}(t+\Delta t)$ using
the  ordinary perturbation calculus. In the first order in $\Delta t$
one obtains then precisely eq.~(\ref{eqn:runU}). This approach also
can serve to justify eq.~(\ref{eqn:crosscond}).
  
Finally   we comment on  two  flavour mixing (which is
sometimes good   approximation  to the more complex   three  flavour
mixing). The $2\times2$ MNS matrix is an ordinary orthogonal rotation matrix.
The general formalism applied to  the mixing between say,
second and third generation, gives 
\begin{eqnarray}
{d\over dt}\sin\vartheta =  \sin\vartheta\cos^2\vartheta
         {C_3+C_2\over C_3-C_2}(h^2_3-h^2_2). 
\label{eqn:2genor}
\end{eqnarray}
It is then straighforward to write down the equation for the evolution
of $\sin^22\vartheta$: 
\begin{eqnarray}
{d\over dt}\sin^22\vartheta = 2 \sin^22\vartheta \cos2\vartheta
           {C_3+C_2\over C_3-C_2}(h^2_3-h^2_2). 
\label{eqn:2gen}
\end{eqnarray}
Observe    also   that
eq.~(\ref{eqn:2gen}), contrary to the original one (\ref{eqn:2genor}),
cannot  be used for   the initial condition $\cos2\vartheta=0$.   This is
because  $\cos2\vartheta=0$ is the point  at  which the uniqueness of the
solution  of the differential   equation (\ref{eqn:2gen}) is violated.
It has  there two solutions: a  t-dependent one which  is the solution
also    to    eq.~(\ref{eqn:2genor})      and    the    second    one,
$\sin^22\vartheta\equiv1$,  which  does not solve eq.~(\ref{eqn:2genor}).
Thus, the previous claims that the maximal mixing $\sin^22\theta=1$ is
stable against the RG running are not correct
\cite{BALEPA,TA,ELLELONA,CAELLOWA,LO}. 
It  is easy to see that,  for the two-flavour   evolution to be  good
approximation   to the full three-flavour  evolution  of $\theta_2$ as
given by  eq.   (\ref{eqn:runs2}),  two  conditions must be    met: i)
$\theta_3$ has  to be small and  ii) the factors $A_{31}$ and $A_{32}$
have to be comparable. 

\section{Evolution of mixing angles}

The equations  derived in  the previous section can be used for a
systematic  analysis  of the RG evolution 
  of the  mass  and  mixing patterns  which are
consistent  with neutrino experimental data.  To fix  the framework we
universally   impose  the following constraints at the electroweak scale:
  $\Delta M^2=10^{-3}$
eV$^2$, $\sin^22\theta_2>0.82$  and  $\sin^2\theta_3\equiv s^2_3<0.2$.
Furthemore,  for each solution -  VO, LAMSW and  SAMSW  - to the solar
neutrino deficit  problem   we  assume  the   corresponding value   of
$\sin^22\theta_1$  and  consider   all  mass  configurations  that are
consistent with  the respective $\Delta m^2$. The  list  of acceptable
low energy mass patterns is, therefore, as follows: 
\begin{itemize}

\item[I] $\Delta M^2\approx m^2_3\gg m^2_1,m^2_2$  (hierarchical)
         with two distinct subpatterns:

  \begin{itemize}

  \item[] $\Delta m^2\approx m^2_{2(1)}\gg m^2_{1(2)}$ or

  \item[] $m^2_{2(1)}\simgt m^2_{1(2)}$ with $m^2_{2(1)}$ ranging from 
      $m^2_{2(1)}\simgt\Delta m^2$ to $m^2_{2(1)}\gg\Delta m^2$ (the 
      latter being, strictly speaking, possible only in the VO case; 
      for SAMSW and LAMSW solutions, with our representative numbers 
      $\Delta m^2=5\times 10^{-6}$ and $5\times10^{-5}$ eV$^2$, respectively,
      $m^2_{2(1)}\gg\Delta m^2$ violates the inequality $\Delta M^2\approx 
      m^2_3\gg m^2_1, m^2_2$ and one gets
      the pattern III below). It is also necessary to distinguish the 
      cases $m_1m_2<0$ and $m_1m_2>0$ \cite{CAESIBNA_VO,BAROST,MA}

  \end{itemize}

\item[II] $\Delta M^2\approx m^2_2,m^2_1\gg m^2_3$ (inversely
          hierarchical). Here, also, the evolution depends on the choice
          $m_1m_2<0$ (with $\Delta m^2>0$ or $<0$) or $m_1m_2>0$ (with 
          $\Delta m^2>0$ or $<0$)

\item[III] $m^2_3\approx m^2_2\approx m^2_1$ and all of the order or larger 
           than $\Delta M^2$ (degenerate). Here the evolution depends on the
           relative magnitudes (four different orderings satisfy our general
           constraints) and relative signs of the masses (we choose $m_3>0$
           and for each ordering there are four inequivalent sign 
           combinations).

\end{itemize}

Since the evolution of angles is less dependent on the evolution of mass 
eigenvalues than the other way around, it is convenient to begin our
discussion with the former. It can be classified into several universal
types of behaviour, depending on the magnitude of the factors $A_{ab}$ 
in eqs.~(\ref{eqn:runs1}-\ref{eqn:runs3}). We note that, neglecting the
effects of mass evolution, all possible mass configurations contained
in patterns I$-$III give one of the following four structures:
\begin{itemize}

\item[a)] $A_{31}\approx A_{32}$ and $|A_{31}|\approx |A_{21}|\approx1$

\item[b)] $A_{31}\approx A_{32}$ and $|A_{21}|\gg|A_{31}|$, $|A_{21}|\gg1$

\item[c)] $A_{32}\approx A_{21}\approx0$, $|A_{31}|\gg1$

\item[d)] $A_{31}\approx A_{21}\approx0$, $|A_{32}|\gg1$

\end{itemize}

For $a)$ and $b)$, since $A_{31}\approx A_{32}$,
equations~(\ref{eqn:runs1}-\ref{eqn:runs3}) reduce to: 
\begin{eqnarray}
\dot s_1&=& -c_1(s_1s_2-c_1c_2s_3)(-c_1s_2-s_1c_2s_3)A_{21}h^2_\tau   -
         c_1s_2c_2s_3A_{32}h^2_\tau,\nonumber\\                    
\dot s_2&=&s_2c_2^2A_{32}h^2_\tau,\label{eqn:runssred}\\   
\dot s_3&=&c^2_2s_3c_3^2A_{32}h^2_\tau.\nonumber 
\end{eqnarray}
Thus,  the evolution  of  the atmospheric mixing  angle  is as in  the
two-generation  equations (\ref{eqn:2genor}) or
(\ref{eqn:2gen}). The first two patterns,  hierarchical and inversely 
hierarchical give the $A_{ab}$ factors as in $a)$ or $b)$. In addition, since 
for patterns I and II $|A_{32}|\approx1$, the evolution of  both, 
$s_2$  and $s_3$ is very  weak (note that  $s_2=0$  and/or $s_3=0$ are
fixed points). Denoting 
\begin{eqnarray}
\xi_\tau\equiv\exp\left(-\int_0^t
2A_{32}(t^\prime)h^2_\tau(t^\prime)dt^\prime\right)
\approx\exp\left(-2A_{32}(0)\epsilon\right) 
\end{eqnarray}
the solution for $s^2_2(t)$ reads 
\begin{eqnarray}
s^2_2(t) = s^2_2(0)/\left[s^2_2(0) + c^2_2(0)\xi_\tau\right] 
\label{eqn:sinatmexact}
\end{eqnarray}
and yields 
\begin{eqnarray}
\sin^22\theta_2(t)  &=&   \xi_\tau\sin^22\theta_2(0)/\left[s^2_2(0)  +
c^2_2(0)\xi_\tau\right]^2                                  \nonumber\\
&\approx&\sin^22\theta_2(0)\left[1+2\cos2\theta_2(0)A_{32}(0)\epsilon\right]
+ {\cal O}(A^2_{32}\epsilon^2). 
\label{eqn:sinatmeps}
\end{eqnarray}
The  effect of the  running
for  $\sin^22\theta_{atm}$ is  a 2.5\% change   for extreme value of
$\tan\beta\approx50$. Since there is
no large factor involved, the logarithmic approximation (the last line
in eq.~(\ref{eqn:sinatmeps})) is very good. This is important for the 
evolution of the masses, discussed in the next section.

Moreover, we observe that, in the very good approximation  of constant
$c_2$,    the      solution  for   $s_3$        is    also   of    the
form~(\ref{eqn:sinatmexact}),  with  $s_2(c_2)\rightarrow  s_3(c_3)$   and
$A_{32}\rightarrow c^2_2A_{32}$ in the definition of $\xi_\tau$.  
 
For the solar mixing angle, in  the approximation $s_3=0$,   we have 
\begin{eqnarray}
\dot s_1=s_1c_1^2s_2^2A_{21}h^2_\tau
\label{eqn:s1fors3zero}
\end{eqnarray}
and the solution for $s_1$ is of the form~(\ref{eqn:sinatmexact}) too,
with $s_2(c_2)\rightarrow s_1(c_1)$ and 
\begin{eqnarray}
\xi_\tau\rightarrow\xi_\tau^\prime=\exp\left(-\int_0^t
2s^2_2(t^\prime)A_{21}(t^\prime)h^2_\tau(t^\prime)dt^\prime\right). 
\label{eqn:xiprime}
\end{eqnarray}
With further specification to the structure $a)$, occuring for 
$\Delta m^2\approx m^2_{2(1)}\gg m^2_{1(2)}$ and, in general, for
$m_1m_2<0$, the evolution of the angle $s_1$ is
very weak, independently of the chosen solar solution. Since all
$|A_{ab}|\simlt1$ the dependence on $s_3$ is weak (for $s_3^2<0.2$).
We conclude that in case $a)$ the evolution of all mixing angles is
negligible and the existence of the fixed points at $U_{ab}=0$ in 
eqs.~(\ref{eqn:runs1}-\ref{eqn:runs3}) is irrelevant.

The evolution of $s_1$ is dramatically different for 
$|A_{21}\epsilon|\gg1$, i.e. for $m^2_{2(1)}>|\Delta m^2|$ and $m_1m_2>0$.
The presence of the fixed points becomes relevant but, since the evolution
of $s_2$ and $s_3$ is still very weak ($|A_{32}|\approx1$), they manifest
themselves as two approximate fixed points of the RG equations for $s_1$. 
One can easily check
(for instance, by considering the equation for 
$d\tan\theta_1/dt=(1/c_1^3)\dot s_1$) that for $A_{21}>0$ (i.e. for
$\Delta m^2>0$) the point $U_{31}=0$ is the UV fixed point and $U_{32}=0$
is the IR fixed point. For $A_{21}<0$ (i.e. for $\Delta m^2<0$) the situation
is reversed. It is also interesting to notice that in the limit $s_3=0$
we can follow analytically the approach to  the fixed points. In this limit
the RG equation for $s_1$ has again the form (\ref{eqn:s1fors3zero}) with the
solution of the form (\ref{eqn:sinatmexact}) with the replacement 
(\ref{eqn:xiprime}). Therefore, for $A_{21}>0$, in the top-down running the 
factor $\xi^\prime\rightarrow0$ exponentially with growing 
$A_{21}h^2_\tau\log(M/M_Z)$ and, consequently, we obtain $s_1(t)=\pm1$ 
(depending on its initial sign) and approach IR fixed point at $U_{32}=0$.
In the bottom-up evolution we approach
$s^2_1(t)\approx0$ exponentially, i.e. the UV fixed point at $U_{31}=0$. 
For $A_{21}<0$ we get the reversed situation, in accord with our general
expectations.

\begin{figure}
\psfig{figure=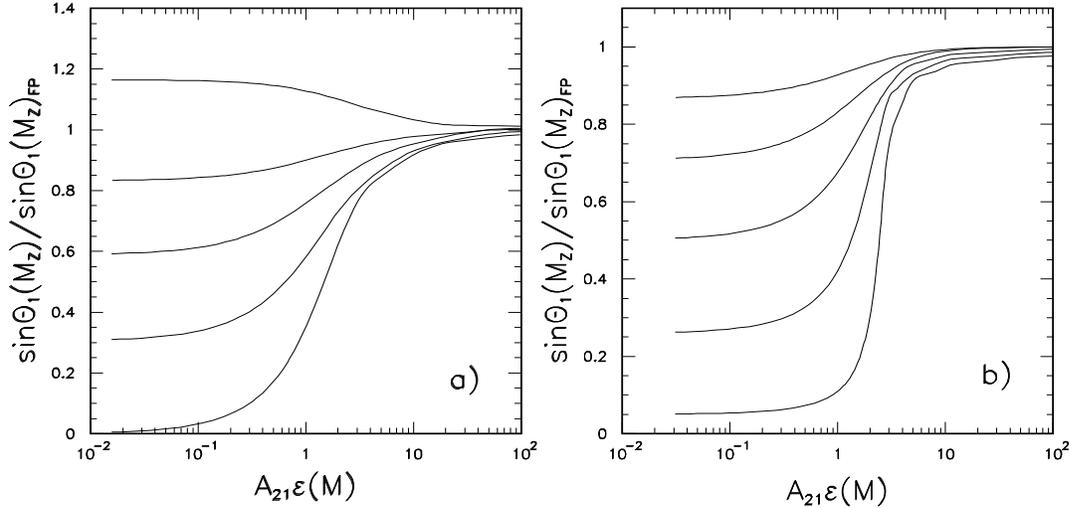,width=15.0cm,height=7.50cm} \vspace{1.0truecm}
\caption{Approach to the IR fixed point for the angle $\theta_1$ 
as a function of $A_{21}\epsilon(M)$
for two different values of $s_3$: a) $s^2_3=0.2$ and the hierarchy
$|m_3|\gg |m_2|\simgt|m_1|$, b) $s^2_3=(0.025)^2$ and the hierarchy
$|m_2|\simgt|m_1|\gg |m_3|$. In both cases $m_1m_2>0$ and we have taken 
$\sin^22\theta_2=0.9$ so that in case a) fixed point corresponds to
$\sin^22\theta_1=0.8$ and in case b) to 
$\sin^22\theta_1=5\times10^{-3}$. Different lines
correspond to different choices of $\theta_1(M)$ As explained in the text, 
for $|A_{21}|\epsilon\gg1$ and $|A_{32}|\approx|A_{31}|\simlt1$, the
evolution of $\theta_2$ and  $\theta_3$ is negligible.}
\label{fig:neu1}
\end{figure}
\vskip 0.3cm

Several comments are in order here. First, it is useful to remember the 
relation between $\sin^22\theta_1$ and $\sin^22\theta_2$ as the function 
of $s_3$ at $U_{31}=0$ and $U_{32}=0$ (i.e. for $\tan\theta_1=s_3/\tan\theta_2$
and $\tan\theta_1=-\tan\theta_2/s_3$, respectively). For both we get 
the following:
\begin{eqnarray}
\sin^22\theta_1 ={s^2_3\sin^22\theta_2\over(s_2^2c^2_3 + s^2_3)^2}
 ~~~~{\rm with} 
 ~~~~s_2^2 = {1\over2}\left(1\pm\sqrt{1-\sin^22\theta_{atm}}\right). 
\label{eqn:mainrel}
\end{eqnarray}
Thus, both fixed points at IR are at present experimentally acceptable with 
the mixing close to bi-maximal for $s^2_3\approx0.2$ and with the SAMSW
solution for $s_3\approx0$. This is interesting in the sense of the
independence of initial conditions at the scale $M$. 

Secondly, it is interesting to estimate the values of $A_{21}$ and $\tan\beta$,
for which the approach to the IR fixed points is seen. This is shown in 
Fig.~\ref{fig:neu1}. We can estimate
that for approaching the fixed point during the evolution in the range 
$(M_Z,M)$ one needs $A_{21}\epsilon(M)>3$, i.e. for $\tan\beta=20$ one needs
$m_1\approx m_2\simgt10^{-4}$ eV for the VO and $m_1\approx m_2\simgt0.01$ eV
for LAMSW or SAMSW solution.

Finally we note, that from the point of view of the initial conditions
at the scale $M$, the UV fixed point looks not realistic as the neglected
muon Yukawa coupling $h_\mu$ quickly destabilizes it during the evolution.
We conclude that for hierarchical and inversely hierarchical mass patterns
the evolution of the mixing angles is either very mild or shows (for
$|A_{21}|\epsilon\gg1$) fixed point behaviour. With higher precision
experiments, it will be very interesting to confirm or disprove the IR
fixed point relation between the angles.

The evolution of mixing angles in the degenerate case, 
$m_3^2\approx m_2^2\approx m_1^2\sim{\cal O}(\Delta M^2)$ or larger, 
partly falls  into the same
classes of behaviour. Indeed, as long as $A_{31}\approx A_{32}$ and
$|A_{21}|\gg|A_{31}|$,
$|A_{21}|\gg1$, the angles evolve according to the same equations
(\ref{eqn:runssred}). One can easily identify the eight mass patterns of the 
degenerate case that fall into this cathegory: the sufficient condition
is that $m_1$ and $m_2$ are of the same sign. For the evolutions of
$s_1$ we then closely follow the two possibilities (depending on the
sign of $A_{21}$) discussed for the first two hierarchies with $m_1m_2>0$. 
We simply note that larger values of $|A_{21}|$ are generic for the present 
case and, as seen in Fig.~\ref{fig:neu1}, the approach to 
the fixed points is faster.
However, the evolution of $s_2$ and $s_3$ are mild only if $m_1$ and 
$m_2$ are negative ($A_{31}\approx A_{32}\approx0$). For positive $m_1$ and 
$m_2$, we have $A_{31}\epsilon\approx A_{32}\epsilon$ and
$|A_{32}\epsilon|$ can be  much larger than 1
(depending on the overall mass scale $0.1 ~{\rm eV}\simlt m_1 \simlt 2$
eV) so that the evolution of $s_2$  and $s_3$ is no longer negligible.
According to  the solution
(\ref{eqn:sinatmexact}), $s_2$  is exponentially  focused to $s_2(t)=0$
or  $s_2(t)=\pm1$, depending on the sign of $A_{32}$, and on the direction
of the evolution but independently of the values of $s_1$ and $s_3$. We  
stress   again that $s_2(0)=\pm1/\sqrt2$ i.e. $\sin^22\theta_{atm}=1$
is  not stable. The angle $s_3$ behaves in a similar way. 
Both angles can approach the value for which $\sin^22\theta_i=0$ ($i=2,3$),
corresponding to the fixed points of their respective equations. This 
is experimentally acceptable  as IR and UV fixed point for $s_3$ but only 
as UV for $s_2$. Since UV fixed points are unstable and small corrections 
push the  evolution towards the IR ones, we conclude that the pattern
$A_{31}\approx A_{32}$, $|A_{31}|\gg1$ is unacceptable in the regime in 
which the approach to the fixed points is relevant.

\begin{figure}
\psfig{figure=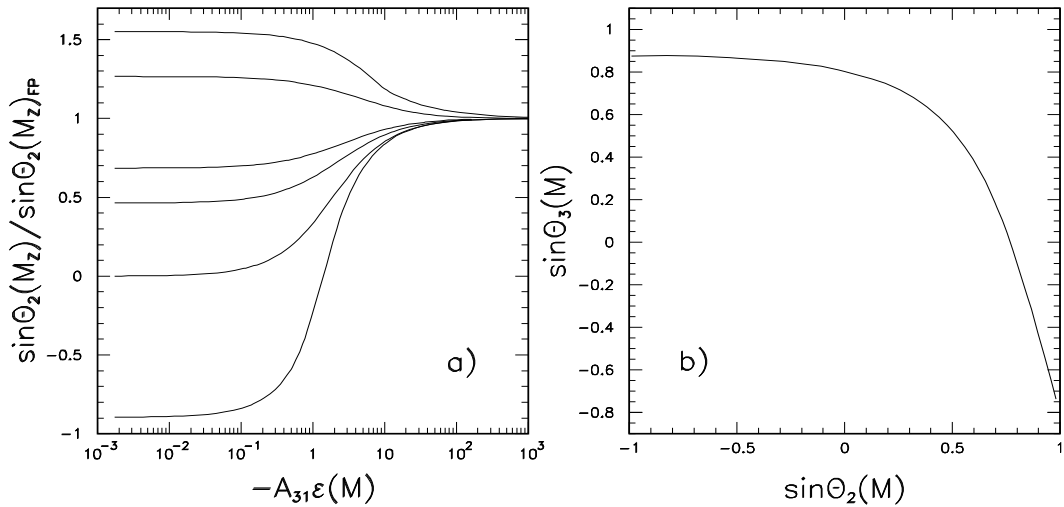,width=15.0cm,height=7.50cm} \vspace{1.0truecm}
\caption{{\bf a)} Approach to the IR fixed point for the degenerate mass 
pattern $|m_2|\approx |m_1|>|m_3|$ and $m_1>0$, $m_2<0$, as a function of 
$-A_{31}\epsilon(M)$ for initial values of $s_2(M)$, $s_3(M)$ and the masses
such that $\Delta m^2(M_Z)=5\times10^{-5}$ eV$^2$, 
$\Delta M^2(M_Z)=10^{-3}$ eV$^2$, $s^2_3(M_Z)=0.2$ and
$\sin^22\theta_1(M_Z)=0.85$ (LAMSW); at the fixed point 
$\sin^22\theta_2(M_Z)=0.9$. Different 
lines correspond to different choices of $\theta_2(M)$
{\bf b)} Correlation of $s_2(M)$ and $s_3(M)$ 
which give experimentally acceptable angles at the fixed point. 
Parameters at $M_Z$ are as in panel a).}
\label{fig:neu2}
\end{figure}

The remaining degenerate mass patterns can be classified according to the
relations $A_{32}\approx A_{21}\approx0$ or $A_{31}\approx A_{21}\approx0$.
Consider first $A_{21}\approx A_{32}\approx0$. The 
equations governing the evolution of the mixing angles can be approximated as
\begin{eqnarray}
\dot s_1 &=& -s_1c_1c_2s_3(s_1s_2-c_1c_2s_3)A_{31}h^2_\tau,\nonumber\\
\dot s_2 &=& s_1c^2_2(s_1s_2-c_1c_2s_3)A_{31}h^2_\tau,\label{eqn:reda}\\
\dot s_3 &=& -c_1c_2c^2_3(s_1s_2-c_1c_2s_3)A_{31}h^2_\tau.\nonumber
\end{eqnarray}
These equations exhibit IR quasi-fixed point behaviour for $A_{31}<0$ 
corresponding to $U_{31}=0$. As before, at the fixed point the angles 
satisfy the relation $s_3 = \tan\theta_1\tan\theta_2$.
We recall that this relation is consistent with present experimental 
information. Since $\dot s_1$ is proportional to $s_1$ and suppressed
by $s_3$, the running of $s_1$ is weak. The IR fixed point is reached due to
strong running of $s_2$ and $s_3$. The rate of approaching the IR fixed 
point by $s_2$ is illustrated in Fig.~\ref{fig:neu2}a. 
For $A_{31}\epsilon\gg1$, $s_2(M_Z)$ is strongly focused at $\pm1$. 
Thus, mass and $\tan\beta$ configurations 
leading to $A_{31}\epsilon\gg1$ are unacceptable.
For $A_{32}<0$ we get IR fixed point in $U_{32}=0$.


\vskip 0.3cm

In summary, with all $|A_{ab}|\simlt1$ the evolution of the mixings is 
negligible. For $|A_{21}|\gg1$ and $|A_{31}|, ~|A_{32}|\simlt1$, or for
$A_{32}(A_{31})\ll-1$ and $A_{31}(A_{32}) \approx A_{21}\approx0$ the 
infrared fixed points are reached during the evolution, independently 
of further details of the mass matrices. 
We also note that for $|A_{21}|\gg1$ only $s_1$ runs to assure the fixed 
point relation, so already the initial values for $s_2$ and $s_3$  at the 
scale $M$ have to be close to their experimental values. For $A_{31}\ll-1$
or $A_{32}\ll-1$, $s_2$ and $s_3$ evolve strongly and the evolution of 
$s_1$ is weak. To assure  consistency with experimental data,
the initial value of $s_1$ has to be close to the experimental value and the 
initial  values of the other two angles have to satisfy certain relation 
(see Fig.~\ref{fig:neu2}b). 
Special examples of the IR fixed point behaviour are the textures 
with exact degeneracy of some masses, which assure the IR fixed point 
relations already at the scale $M$. They will be recalled in the last part 
of the paper. 

\section{Evolution of mass eigenvalues}

The guiding principle for understanding the effects of evolution of
mass eigenvalues is eq.~(\ref{eqn:runmassdiff}) and the evolution of the 
angles. It follows from eq.~(\ref{eqn:runmassdiff}) that quantum corrections
$|\Delta m^2_{ab}(M_Z)-\Delta m^2_{ab}(M)|$ are limited from above by 
${\cal O}(m^2_{a(b)}\epsilon)$ where 
$\epsilon\approx \tan^2\beta\times10^{-5}$.
In fact, generically this upper bound is saturated (in the sense of order
of magnitude) for those mass patterns which give fixed point solutions
for mixing angles (at all scales or reachable after the evolution) since
then in general $\langle U^2_{31}\rangle\neq \langle U^2_{32}\rangle$
(where $\langle\dots\rangle$ means average over the evolution). 
Quantum corrections orders of 
magnitude smaller than this upper bound can be naturally guaranteed only
for weakly evolving angles and with $U^2_{31}\approx U^2_{32}$. Those two
rules determine the order of magnitude of quantum corrections to any
mass pattern of interest. We illustrate them with several examples.

The fully hierarchical pattern, with 
$\Delta M^2\approx m^2_3\gg|\Delta m^2|
\approx m^2_{1(2)}\gg m^2_{2(1)}$, is obviously not altered by quantum
corrections. Similar conclusion holds for 
$m^2_1\approx m^2_2\sim{\cal O}(\Delta m^2)$. More interesting effects may 
appear for the hierarchy 
$\Delta M^2\approx m^2_3\gg m^2_1,m^2_2\gg \Delta m^2$ which is possible
for the VO solution. It is then 
interesting to ask the same question which is usually asked 
\cite{CAESIBNA_VO,CAESIBNA_SM,CAESIBNA_SUSY,BAROST,ELLO}
for the inversely hierarchical and degenerate patterns: is the initial
condition $m_1=\pm m_2$ at the scale $M$ compatible with the measured
(at the scale $M_Z$) $\Delta m^2$ ? We consider first $m_1=m_2$ which
places us in the structure $b)$
considered previously for the evolution of the angles.  We note 
that eq.~(\ref{eqn:crosscond}) imposes (for $h_\mu=h_e=0$) at the scale $M$
the fixed point values $U_{31}=0$ or $U_{32}=0$. It follows then from the 
evolution of the masses, eq.~(\ref{eqn:solmass}) and our discussion of the 
fixed point solutions, that both choices are IR fixed points, so remain 
stable during the evolution. For $U_{31(32)}=0$ we get 
$\Delta m^2_{21}(M_Z)=\pm m^2_{1(2)}(M) 4\epsilon U^2_{32(31)}$,
where $U^2_{32(31)}\sim{\cal O}(s_2^2)$. So, for 
$m^2_1=m^2_2\sim10^{-(6-8)}$ eV and $\tan\beta\simlt2.5$ one obtains the right 
order of magnitude for the VO solution \cite{CAESIBNA_VO}. 

Choosing $m_1=-m_2$, the angles remain uncorrelated with the masses (structure
$a)$) and stable. For $m^2_{1(2)}\sim10^{-(5-8)}$ eV$^2$ it is easy to choose 
angles so that  $\Delta m^2_2(M_Z) \sim 
{\cal O}(\epsilon m^2_{1(2)})\sim10^{-10}$ eV$^2$
in a large range of $\tan\beta$ values.
 
One should also note that for the 
hierarchy I with $m_1^2(M) = m^2_2(M)$ getting the right $\Delta m^2$ for 
SAMSW or LAMSW solutions is only marginally possible for $m_1^2\simgt10^{-4}$
eV$^2$ and large values of $\tan\beta$ ($>50$).

We consider now the inverse hierarchy 
$\Delta M^2\sim m_1^2\approx m^2_2\gg m^2_3$. For $m_1(M)=m_2(M)$ we can 
repeat the previous discussion remembering, however, that now 
$m^2_{1(2)}\sim10^{-3}$ eV$^2$. Thus, with such initial conditions we cannot
reproduce $\Delta m^2$ of the VO solution but we naturally get $\Delta m^2$ 
of the SAMSW (for $2.5\simlt\tan\beta\simlt25$)
or LAMSW (for $5\simlt\tan\beta$)
solutions \cite{CAESIBNA_VO}. And, of course, the mixing 
angles satisfy  one of the fixed point relations for all scales. 
Note, however, that as stressed earlier the exact degeneracy at $M$ is not
necessary to reach the IR fixed point at $M_Z$

The choice $m_1=-m_2$ guarantees mild evolution of the elements $U_{3a}$,
$U_{3a}(t)=U_{3a}(0)+{\cal O}(\epsilon)$. Since $m^2_1\approx m^2_2\sim10^{-3}$
eV$^2$  and $\Delta m^2_{21}(M_Z)\simlt{\cal O}(\epsilon m^2_{1(2)}(M))$ we
note that choosing the angles in agreement with experimental information we 
can reproduce the right order of magnitude for $\Delta m^2$ for the LAMSW or
SAMSW solutions  (for roughly the same ranges of $\tan\beta$ as for
$m_1=m_2$). On the other hand, for the VO solution, one needs 
$\Delta m^2_{21}(M_Z)\approx{\cal O}(\epsilon^2 m^2_{1(2)}(M))$ \cite{BAROST}.
The absence of ${\cal O}(\epsilon m^2_{1(2)}(M))$ contribution to $\Delta m^2$ 
requires $U^2_{31}=U^2_{32}$ which has two solutions:
\begin{eqnarray}
s_3 = -\left({s_2\over c_2}\right){c_1-s_1\over c_1+s_1} \equiv 
{s_2\over c_2}\tan\left(\theta_1-{\pi\over4}\right)~~~{\rm or}
\label{eqn:correlation1}
\end{eqnarray}
\begin{eqnarray}
s_3 = \left({s_2\over c_2}\right){c_1+s_1\over c_1-s_1} \equiv 
-{s_2\over c_2}\cot\left(\theta_1-{\pi\over4}\right)\phantom{aa}
\label{eqn:correlation2}
\end{eqnarray}
independently of $\tan\beta$.

The  easiest way to   find
consistently the mass splitting in order  $\epsilon^2$ is to solve the
coupled  equations (\ref{eqn:runC},\ref{eqn:runs1}-\ref{eqn:runs3}) in
order $\epsilon^2$ by using the so-called Banach's  principle i.e.  by
substituting  in the RHSs of the  RGEs their solutions obtained in the
order $\epsilon$ and integrating once more. One then gets (for definiteness 
we assume~(\ref{eqn:correlation1}) and following ref. \cite{BAROST} take the 
neutrino mass spectrum in the form $\sim(-1,1,z)$): 
\begin{eqnarray}
m_1^2 &=& m^2_\nu\left(1-4U^2_{31}\epsilon+8U^4_{31}\epsilon^2 
-4U^2_{31}c_2^2c_3^2{z-1\over z+1}\epsilon^2\right),\nonumber\\
m_2^2 &=& m^2_\nu\left(1-4U^2_{31}\epsilon+8U^4_{31}\epsilon^2  
-4U^2_{31}c_2^2c_3^2{z+1\over z-1}\epsilon^2\right).
\end{eqnarray}
where $U^2_{31}=U^2_{32}= (s_2/(s_1+c_1))^2$.
We have used the approximation of constant  $h_\tau$ and absorbed the
overall   renormalization  introduced    by   the   factor   $K$    in
eq. (\ref{eqn:runC})  in $m_\nu$.  We obtain, therefore 
\begin{eqnarray}
\Delta m^2 = -16 m^2_\nu s_2^2c^2_2 \left({c_3\over c_1+s_1}\right)^2
{z\over z^2-1}\epsilon^2
\end{eqnarray}
For $s_3=0$ and $s_1^2=c_1^2=1/2$ this coincides with  the result  derived 
in ref. \cite{BAROST}.
Since $\Delta M^2=m^2 (z^2 - 1) + {\cal O}(\epsilon)$ we get \cite{BAROST}
\begin{eqnarray}
\Delta m^2 = -16 \left({m^4_\nu\over\Delta M^2}\right)
s_2^2c^2_2 \left({c_3\over c_1+s_1}\right)^2 z \epsilon^2
\end{eqnarray}
We note that for inversely hierarchical pattern, $z=0$,
${\cal O}(\epsilon^2)$ contribution to $\Delta m^2$ vanishes too. 
For $z\sim{\cal O}(1)$ the above considerations are applicable also to the
degenerate pattern III provided $|A_{31(2)}|\epsilon\simlt1$ (see 
Fig.~\ref{fig:neu2}a). With $\epsilon<2.5\times10^{-2}$ for $\tan\beta<50$
we get $A_{31(2)}<50$ and $z<49/51$. For a large range of parameters it is 
easy to obtain $\Delta m^2$ consistent with the VO solution.
For the degenerate pattern III with $A_{31(2)}\epsilon\ll-1$  
the expansion in $\epsilon$ (in fact for the 
mixing angles this is an expansion in $\epsilon A_{32}$) breaks down and
this mechanism cannot work.

In general, with the degenerate pattern III, the potentially  
relevant simple textures are 
\begin{eqnarray}
m_3> |m_1|=|m_2|  ~~~~~~{\rm or}  ~~~~~~|m_1|=|m_2|>m_3.\nonumber
\end{eqnarray}
The exact degeneracy of absolute values $|m_3|=|m_2|=|m_1|$ would not 
result in two distinctly different $\Delta M^2$ and $\Delta m^2$ after 
evolution.
With $m_1$ and $m_2$ both  negative we are in the IR fixed points and
encounter similar situation as for the inversely hierarchical case
with $m_1=m_2$. The result
$\Delta m^2_{21}(M_Z)\sim{\cal O}(\epsilon m^2_{1(2)}(M))$ is compatible
with the LAMSW and SAMSW solutions but not with the VO
solution. However, the relation
$\Delta m^2_{31}(M_Z)\sim\Delta m^2_{32}(M_Z)\approx\Delta M^2$ can be 
simultaneously satisfied only if all masses squared are of order 
$\Delta M^2\sim10^{-(3-2)}$
and therefore we need $2.5\simlt\tan\beta\simlt15$ 
($5\simlt\tan\beta$) for 
the SAMSW (LAMSW) solution.
As explained in the previous section, both masses positive are unacceptable 
in the regime in which the fixed points are relevant.

For $m_1=-m_2$ we get $A_{21}=0$, $A_{32}\approx0$ for positive $m_1$
and $A_{21}=0$, $A_{31}\approx0$ for negative $m_1$, with $|A_{31}|\gg1$
and $|A_{32}|\gg1$, respectively. Such initial conditions assure the 
reaching of IR fixed points at $U_{31}=0$ for $A_{31}\epsilon\ll-1$ or 
$U_{32}=0$  for $A_{32}\epsilon\ll-1$ and, from the point of view of 
quantum corrections to 
the masses, we are back to  the same situation as for $m_1$ and $m_2$ both 
negative. For $|A_{31(2)}|\epsilon\simlt1$ but $A_{31(2)}\ll-1$ the angles
evolve very mildly and the previously described mechanism of
stabilization can assure right $\Delta m^2$ for the VO solution

\section{Conclusions}

In this paper we have derived RG equations directly for the mixing angles 
and mass eigenvalues. These equations allow for easy qualitative discussion
of the evolution of masses and mixing angles, which systematizes existing
results \cite{ELLO,CAESIBNA_VO,BAROST,CAESIBNA_SM,CAESIBNA_SUSY,HASU} and
allows for their generalization.
The equations for the mixing angles have IR fixed points
at $U_{31}=0$ or $U_{32}=0$, which are reachable for several mass hierarchies
consistent with the measured mass squared differences and characterized by 
one large factor $|A_{ab}|\equiv|(m_a+m_b)/(m_a-m_b)|\gg1$ with all other 
$|A_{ab}|$'s $\simlt{\cal O}(1)$.
Both fixed points give relation (\ref{eqn:mainrel})
which is at present acceptable experimentally.
Its further verification, by clarifying the LSND puzzle, 
measuring $s_3$ and discriminating
between large and small solar angle solutions is of obvious interest.

Special examples of the IR fixed points behaviour are the textures with 
exact degeneracy of some masses. The mass squared differences consistent with 
the measured values can be obtained for the following textures:
{\sl i)} hierarchical with $|m_3|\sim10^{-1.5}$ eV, 
$m_1= m_2$ and $|m_1|\sim10^{-(3-4)}$ eV (for VO),
{\sl ii)} inversely hierarchical with 
$m_1= m_2\sim10^{-1.5} ~{\rm eV} ~\gg m_3$ (for LAMSW and SAMSW),
{\sl iii)} degenerate with $|m_i|\simgt{\cal O}(0.1)$ eV ($i=1,2,3$)
and $m_1=m_2<0$ or  $m_1=-m_2$ with $A_{31}\epsilon\ll-1$ 
(for LAMSW and SAMSW).

With the fully hierarchical masses $|m_3|\sim10^{-1.5}$ eV,
$m_2^2\sim\Delta m^2\gg m_1^2$ and for the inverse hierarchy with
$m_1=-m_2$ the evolution of the angles is very mild. The proper
order of magnitude for $\Delta m^2$ is possible for 
the LAMSW and SAMSW solutions. Similarly mild evolution occurs
for the degenerate case with $m_1=-m_2$, $|A_{31(2)}|\epsilon\simlt1$,
$A_{31(2)}\ll-1$ which is consistent with the stabilization 
mechanism leading to $\Delta m^2$ of the VO solution.

We focused   our discussion on the MSSM. The corresponding RG
equation in the SM are obtainable by the replacements mentioned in the text.
The UV and IR fixed points are interchanged and the running of the 
angles is weaker by factor 2 than for $\tan\beta=1$ in the MSSM. 

\vskip 0.5cm

\noindent {\bf Acknowledgments}
\vskip 0.3cm
\noindent The work of P.H.Ch. was partly supported by the Polish State
Committee for Scientific Research grant 2 P03B 030 14 (for 1998--1999)
and by  the  Maria-Sk\l odowska-Curie Joint Fund  II (MEN/DOE-96-264).
The work of W.K.  and S.P.  was partly supported  by the Polish  State
Committee  for  Scientific    Research grant   2    P03B 052  16  (for
1999--2000).  P.Ch. and S.P. would like  to thank Santa Cruz Institute
for Particle Physics  and  Institute de  Physique Nucleaire in   Lyon,
respectively, for  warm  hospitality  during  the  completion  of this
work.   S.P.  is  indebted   to  Stavros  Katsanevas for   interesting
discussions on neutrino physics


\begin{thebibliography}{99}

\bibitem{ATMOSPH} Y. Fukuda et al., Superkamiokande Collaboration, 
                  {\sl Phys. Lett.} {\bf B433} (1998) 9, {\sl Phys. Rev. Lett.}
                  {\bf 81} (1998) 1562; S. Hatakeyama et al.,
                  Kamiokande Collaboration, {\sl Phys. Rev. Lett.} {\bf 81}
                  (1998) 2016.

\bibitem{SOLAR} B.T. Cleveland et al., {\sl Astrophys. J.} {\bf 496} (1998) 
                505;
                K.S. Hirata et al., Kamiokande Collaboration,  
                {\sl Phys. Rev. Lett.} {\bf 77} (1996) 1683;
                W. Hampel et al., GALLEX Collaboration, 
                {\sl Phys. Lett.} {\bf B388} (1996) 384;
                D.N. Abdurashitov et al., SAGE Collaboration,
                {\sl Phys. Rev. Lett.} {\bf 77} (1996) 4708;
                Y. Suzuki, talk at the 
                conference {\sl Neutrinos'98} Takayama, Japan, June 1998.

\bibitem{CHOOZ} A. Apollonio et al., CHOOZ Collaboration,  {\sl Phys. Lett.}
                {\bf B420} (1998) 397.

\bibitem{LSND} B. Achkar et. al., LSND Collaboration {\sl Nucl. Phys.}
               {\bf B434} (1995) 503, D.H. White, talk at the 
               conference {\sl Neutrinos'98} Takayama, Japan, June 1998.

\bibitem{KR} W. Kr\'olikowski preprints IFT-99/08 (hep-ph/9904489) and
             IFT-99/24 (hep-ph/9910308).

\bibitem{MANASA} Z. Maki, M. Nakagawa and S. Sakata, {\sl Prog. Theor. Phys.}
                 {\bf 28} (1962) 870.

\bibitem{BIGIGR} S.M. Bilenkii, C. Giunti and W. Grimus, preprint 
                 UWTHPH-1998-61 (hep-ph/9812360); S.T. Petcov, talk on
                 17$^{th}$ {\sl Int. Workshop on Weak Interactions and 
                 Neutrinos}, Cape Town, South Africa, January 1999. 

\bibitem{BAR} R. Barbieri et al., {\sl JHEP} (1998) 9812:017.

\bibitem{FOLIMASC} G.L. Fogli, E. Lisi, A. Marrone and G. Scioscia,
                   {\sl Phys. Rev.} {\bf D59} (1999) 033001.

\bibitem{KIKI} J.S. Kim and C.W. Kim, hep-ph/9909428


\bibitem{DIHARA} S. Dimopoulos, L.J. Hall and S. Raby, {\sl Phys. Rev.}
                 {\bf D47} (1993) 3697.

\bibitem{BLRATO} T. Blazek, S. Raby and K. Tobe, preprint
                 OHSTPY-HEP-T-98-030 (hep-ph/9903340).

\bibitem{LELORO} G.K. Leontaris, S.Lola and G.G. Ross, {\sl Nucl. Phys.}
                {\bf B454} (1995) 25.

\bibitem{LELOSCH} G.K. Leontaris, S. Lola, C. Scheich, {\sl Phys. Rev.}
                  {\bf D53} (1996) 6381.

\bibitem{LOVE} S. Lola, J.D. Vergados, {\sl Prog. Part. Nucl. Phys.}
               {\bf 40} (1998) 71.

\bibitem{BAHAST} R. Barbieri, L.J. Hall, A. Strumia,
                 {\sl Phys. Lett.} {\bf B445} (1999) 407.

\bibitem{ALFE} G. Altarelli and F. Feruglio,
               {\sl Phys. Lett.} {\bf B439} (1998) 112,
               {\sl JHEP}  {\bf  9811:021} (1998),
               {\sl Phys. Lett.} {\bf B451} (1998) 388,
               preprint CERN-TH-99-129 (hep-ph/9905536), to appear
               in proceedings of {\sl 8th International Workshop on 
               Neutrino Telescopes}, Venice, Italy, Feb 1999;
               G. Altarelli, F. Feruglio and I. Masina,
               preprint CERN-TH-99-147 (hep-ph/9907532).

\bibitem{LORO} S. Lola and G.G. Ross, {\sl Nucl. Phys.} {\bf B553} (1999) 81.

\bibitem{BAHAKARO} R. Barbieri, L.J. Hall, G.L. Kane and G.G. Ross,
                  preprint OUTP-9901-P (hep-ph/9901228).

\bibitem{JESU} M. Je\.zabek and Y. Sumino, {\sl Phys. Lett.} {\bf B440}
               (1998) 327, {\sl Phys. Lett.} {\bf B457} (1999) 139. 

\bibitem{CAELLOWA} M. Carena, J. Ellis, S. Lola and C.E.M. Wagner,
                   preprint CERN-TH-99-173 (hep-ph/9906362).

\bibitem{ST} B. Stech, preprint HD-THEP-99-15 (hep-ph/9905440).

\bibitem{OLPO} M. Olechowski and S. Pokorski, {\sl Phys. Lett.} {\bf B257}
               1991, 388.

\bibitem{BAROST} R. Barbieri, G.G. Ross and A. Strumia, preprint OUTP-99-30-P
                 (hep-ph/9906470).

\bibitem{CAESIBNA_VO} J.A. Casas, J.R. Espinosa, A. Ibarra and I. Navarro,
                      {\sl JHEP} 9909:015 (1999).

\bibitem{HASU} N. Haba, N. Okamura and M. Sugiura, preprint KEK-TH-597
               (hep-ph/9810471); N. Haba, Y. Matsui, N. Okamura and 
               M. Sugiura, preprint KEK-TH-620 (hep-ph/9904292); 
               N. Haba and N. Okamura,  preprint KEK-TH-632 (hep-ph/9906481).

\bibitem{WE} C. Wetterich, {\sl Nucl. Phys.} {\bf B187} (1981) 343.

\bibitem{CHPL} P.H. Chankowski and Z. P\l uciennik, {\sl Phys. Lett.} 
               {\bf B316} (1993) 312.

\bibitem{CH} P.H. Chankowski, {\sl Phys. Rev.} {\bf D41} (1990) 2877.

\bibitem{BALEPA} K. Babu, C.N. Leung and J. Pantaleone, {\sl Phys. Lett.} 
                 {\bf B319} (1993) 191.

\bibitem{ELLO} J. Ellis and S. Lola, {\sl Phys. Lett.} {\bf B458} (1999) 310.

\bibitem{BA} K. Babu, {\sl Z. Phys.} {\bf C35} (1987) 69.

\bibitem{TA} M. Tanimoto, {\sl Phys. Lett.} {\bf B360} (1995) 41.

\bibitem{ELLELONA} J. Ellis, G.K. Leontaris, S. Lola and D.V. Nanopoulos,
                   {\sl Eur. Phys. J.} {\bf C9} (1999) 389-408.

\bibitem{LO} S. Lola, preprint CERN-TH-99-40, to appear in the proceedings 
             of the 6th Hellenic School and Workshop on Elementary Particle 
             Physics, Corfu, Greece, Sep 1998 (hep-ph/9903203).

\bibitem{MA} E. Ma, preprint UCRHEP-T259 (hep-ph/9907400)

\bibitem{CAESIBNA_SM} J.A. Casas, J.R. Espinosa, A. Ibarra and I. Navarro,
                      {\sl Nucl. Phys.} {\bf B556} (1999) 3.

\bibitem{CAESIBNA_SUSY} J.A. Casas, J.R. Espinosa, A. Ibarra and I. Navarro,
                        preprint CERN-TH/99-142 (hep-ph/9905381).


\end{thebibliography}
\end{document}